\documentclass[twocolumn,twoside,slac_two]{revtex4}
\usepackage{graphicx}
\usepackage{fancyhdr}
\begin{document}

\title{Long-term Monitoring of Accreting Pulsars with Fermi GBM}

%

\author{Mark H. Finger}
\affiliation{Universities Space Research Association, 6767 Old Madison Pike, Huntsville, AL 35806}

\author{Elif Beklen}
\affiliation{Physics Department, Middle East Technical University, 06531, Ankara, Turkey}

\author{P. Narayana Bhat}
\author{William S. Paciesas}
\author{Valerie Connaughton}
\affiliation{Physics Department, University of Alabama in Huntsville, Huntville, AL 35805, USA}

\author{David A. H. Buckley}
\affiliation{South African Astronomical Observatory, PO Box 9, Observatory, 7935, South Africa}

\author{Ascension Camero-Arranz}
\affiliation{Fundaci\'{o}n Espa\~{n}ola de Ciencia y Tecnolog\'{i}a, C/Rosario Pino,14-16, 28020- Madrid, Spain}

\author{Malcolm J. Coe}
\affiliation{School of Physics \& Astronomy, The University, Southampton, SO17 1BJ, United Kingdom}

\author{Peter Jenke}
\affiliation{Oak Ridge Associated Universities / NPP, Oak Ridge, TN 37831}

\author{Gottfried Kanbach}
\affiliation{Max-Planck-Institut fuer extraterrestrische Physik,  Postfach 1603, 85740 Garching, Germany }

\author{Ignacio Negueruela}
\affiliation{Departamento de F\'{i}sica, Ingenier\'{i}a de Sistemas y Teor\'{i}a de la Se\~{n}al, 
Escuela Polit\'{e}cnica Superior,  Universidad de Alicante 
Apartado de Correos 99,  E03080 Alicante,  Spain }

\author{Colleen A. Wilson-Hodge}
\affiliation{NASA / Marshall Space Flight Center, Huntsville, Alabama, 35812}

\begin{abstract}
Using the Gamma ray Burst Monitor (GBM) on Fermi we are monitoring accreting pulsar
systems. We use the rates from GBM's 12 NaI detectors in the 8-50 keV range to detect and
monitor pulsations with periods between 0.5 and 1000 seconds.  After discussing our
analysis approach we present results for individual sources from the first year of
monitoring. Updated figures for these and other sources are available at
http://gammaray.nsstc.nasa.gov/gbm/science/pulsars/\,.

\end{abstract}

\maketitle

\thispagestyle{fancy}


\section{Pulsar Monitoring with GBM}
The full sky coverage of GBM enables long term monitoring of
the brighter accreting pulsars, allowing precise measurements of spin
frequencies and orbital parameters, studies of spin-up or spin-down
rates and hence the flow of angular momentum, and detection and study of new
transient sources or new outbursts of known transients.

\section{Data Analysis}
The analysis of GBM observations of pulsars presents two main
challenges: the background rates are normally much larger than the
source rates, and have large variations; and the responses of the detectors 
to a source are continuously changing because of FermiÕs ever changing orientation.  
We meet these challenges with an empirical background subtraction and extensive
on the fly calculations of detector response matrices. 
The initial step in our analysis consists of screening the NaI detector count rate data to 
remove phosphorescence events, gamma-ray burst and particle events and intervals of rapid spacecraft rotation. 
Then an empirical background model is fit to the rates and subtracted off, after which the
rates are combined over detectors in a way that results in an estimate of the variable part of the  source flux.
From these flux variation estimates we conduct searches for pulsations, and monitor detected sources.

\subsection{Empirical Background Subtraction}

\begin{figure}[t!]
\centering
\includegraphics[width=3.2in]{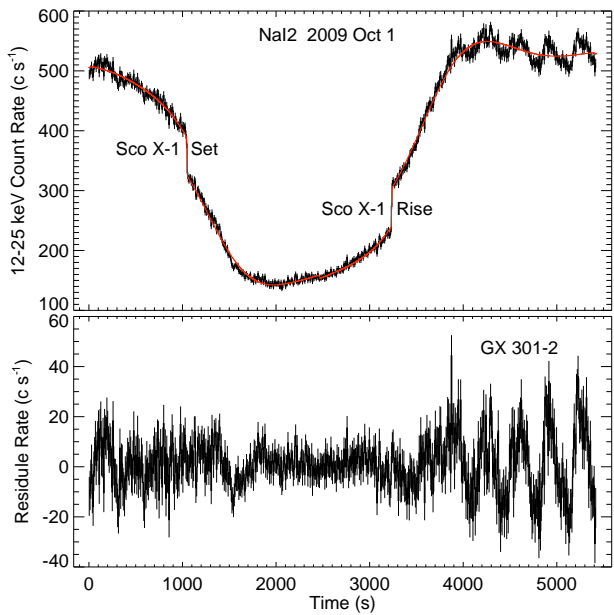}
\caption{A fit of the background rates for NaI detector 2 on 2009 Oct 1 in the 12-25 keV band.
The upper panel show the rates and the fit (red curve), and  the fit residuals are show in bottom panel. } \label{bkg_fit}
\end{figure}

The screened rates in each channel of the 12 NaI detectors are fit with a model with that includes
components for a small number of bright sources, pulse stiff empirical model that contains the low frequency 
component of the remaining rates.  The source models are computed from a model flux spectrum using time dependent 
response matrices. One fit is made for each channel with the normalization factors for the bright sources common 
to all detectors, but with the stiff model parameters (a statistically constrained spline) independent for each detector.
In Figure \ref{bkg_fit} we show an example fit for and residuals for one channel and detector.  The model
successfully fits the occultation steps of Sco X-1, and the overall background trend underlying the pulsation
of GX 301-2 which is evident from the residuals.

\subsection{Estimated Flux Variations}

For a given source we obtain from the NaI detector rate residuals an estimate of the variable part of the source's flux.
Using a model of the source spectrum and the time dependent
detector responses we compute for a given channel the source induced rate $\mu_{ik}$ expected in 
detector $i$ at time $t_k$ if the source 
has unit flux in the channelÕs energy range. The variable part of the flux $\Delta F_k$ is then 
estimated by minimizing  
\begin{equation}
\chi^2_k = \sum_i {{(\Delta r_k -\mu_{ik}\Delta F_k )^2} \over {\sigma^2_k}}
\end{equation}
where  $\Delta r_k$ is the residual rates and $\sigma_k$ their errors.

\subsection{Pulse Searches}

We have implemented two different pulse search strategies, daily blind searches, and source specific searches.

For the daily blind searches we compute fluxes from a days data for 24 source directions equally spaced on the galactic plane. 
For each direction we do an FFT based search from 1 mHz to 2 Hz. These provide us sensitivity to previously unknown sources,
or sources where we have poor estimates of the pulse period. When we detect a source we determine a coarse estimate of the
source's galactic longitude based on the longitude with the highest power. The latitude can also be constrained by mapping
the power over a grid of source directions. 

Source specific searches are made over small ranges of frequency and sometimes frequency rate based on phase shifting 
and summing pulse profiles that are made from short intervals of data, using barycentered and possibly orbitally corrected times.
They provide higher sensitivity than the blind searches because of the smaller search ranges and the use of source specific
information such as the location, orbital parameters and flux spectrum. 

\subsection{Phase Averaged Flux}

\begin{figure}[t!]
\centering
\includegraphics[width=3.2in]{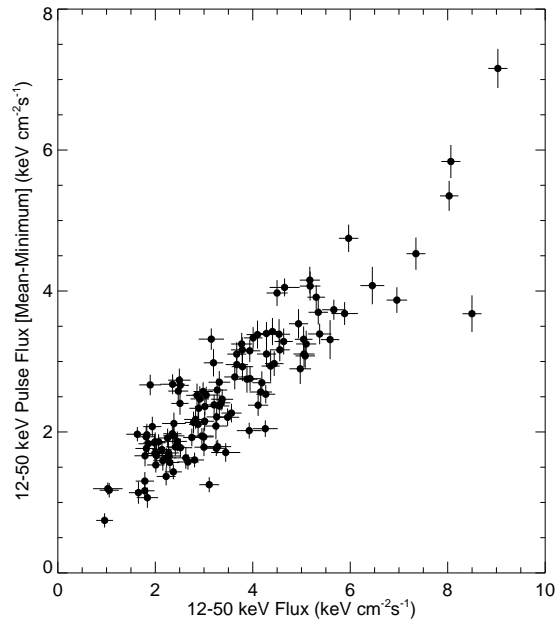}
\caption{The pulsed flux verses the phase averaged flux for Vela X-1 in the 12-50 keV band.} \label{flux_flux}
\end{figure}

Since our background subtraction removes the phase averaged flux, this must be determined from Earth occultation measurements. 
For a dicussion of GBM Earth occultation measurements see C. Wilson-Hodge et al. in these proceedings \cite{Wilson-Hodge_2009}. 
Figure \ref{flux_flux} shows the correlation between pulsed flux and the total flux from Earth occulations for Vela X-1.
Here we show the mean-minimum (or "area") pulsed flux.

\section{Detected Sources}

\begin{figure*}[!t]
\centering
\includegraphics[width=5.5in]{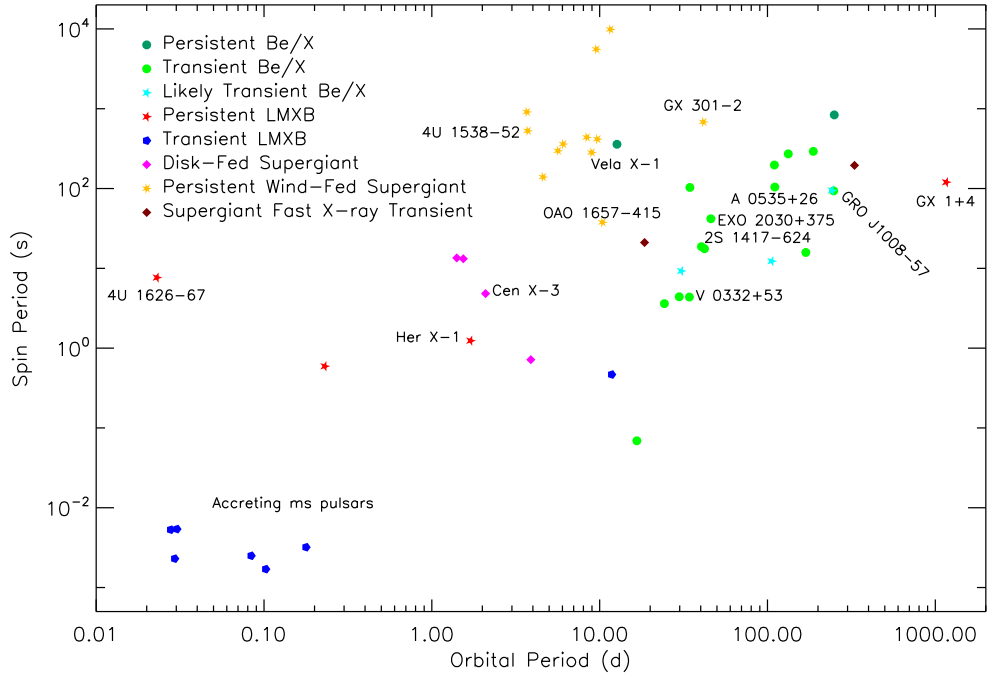}
\caption{The Corbet diagram showing spin period versus orbital period for accreting pulsars. The labeled sources have been detected
by GBM. Four sources with unknown orbital period have also been detected.} \label{corbet_diagram}
\end{figure*}

To date we have detected 17 accreting pulsar systems. In Figure \ref{corbet_diagram} we show how these sources are distributed in the Corbet
diagram, which tends to separate classes of sources. Four of detected sources, 
Cep X-4, A 1118-616, Swift J05131.4-6547 (in the LMC), and MXB 0656-074, are not shown on the figure because their orbital periods are unknown.
These are all transients sources that are known or suspected Be/X-ray pulsar binary systems.

\begin{figure}[!b]
\centering
\includegraphics[width=3.2in]{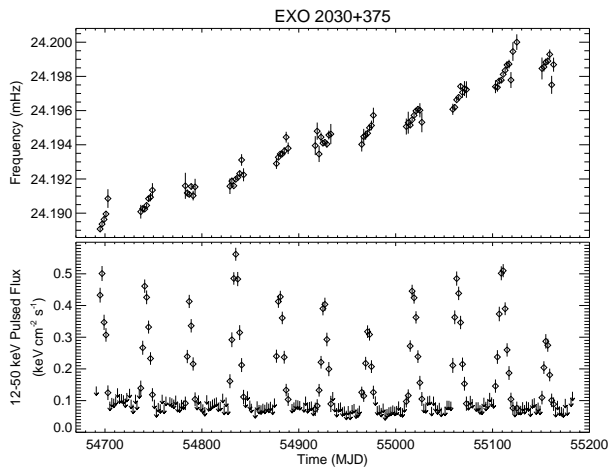}
\caption{Frequency and pulsed flux history for the Be/X-ray binary pulsar EXO 2030+375 ($P_{orbit} = 46.0$\,d, $P_{spin}=41.3$\,s).} \label{exo2030}
\end{figure}

\begin{figure}[!b]
\centering
\includegraphics[width=3.2in]{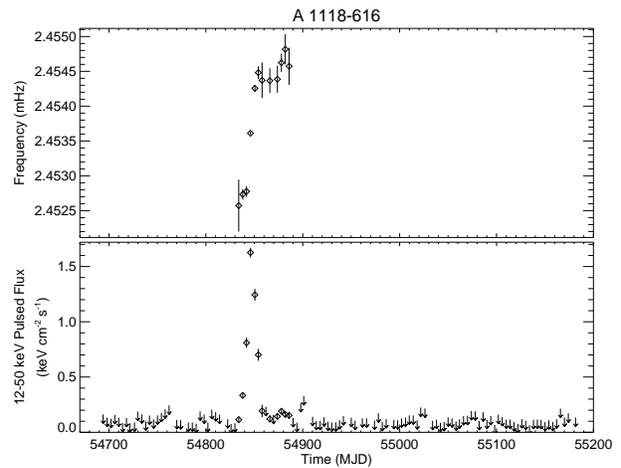}
\caption{Frequency and pulsed flux history for the Be/X-ray binary pulsar A1118-616 ($P_{orbit}$ unknown, $P_{spin}=407.6$\,s).} \label{a1118}
\end{figure}

\subsection{Be/X-ray transients} 

\begin{figure}[!t]
\centering
\includegraphics[width=3.2in]{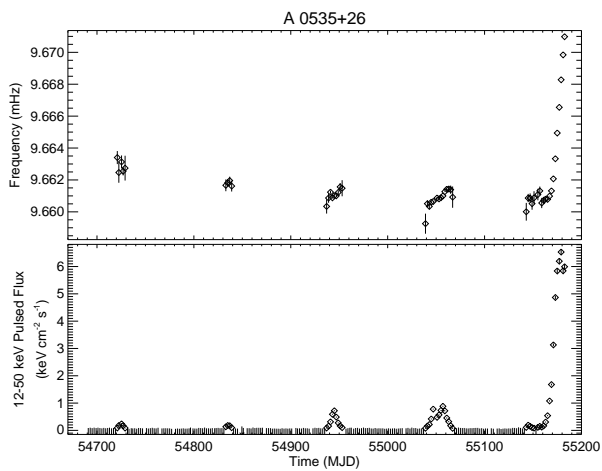}
\caption{Frequency and pulsed flux history for the Be/X-ray binary pulsar A 0535+26 ($P_{orbit} = 111$\,d, $P_{spin}=103$\,s).} \label{a0535}
\end{figure}

We have detected outbursts from the Be/X-ray pulsar systems EXO 2030+375, Cep X-4, V0332+53, A0535+26,
MXB 0656+072, Swift J0513.4-6447 (in the LMC), GRO J1008-57, A1118-615, and 2S 1417-624. These sources show two types of outbursting behavior.
Type I consists of series of several outburst all occurring near periastron passage in their eccentric orbit. These ``normal'' outbursts
typically have luminosities of $10^{36}$ -- $10^{37} {\rm erg~s}^{-1}$. Type II consists of a single ``giant" outburst, often with super-Eddington
peak luminosity, which last for several orbital periods \cite{Stella_White_Rosner}. In Figure \ref{exo2030} we show the GBM spin frequency 
and pulsed flux history of
EXO 2030+375 which shows a good example of the type I behavior. In Figure \ref{a1118} we show the GBM pulse frequency and pulsed flux history
of A 1118-615, which had a single giant outburst (Type II behavior).  Finally in Figure \ref{a0535} we show the GBM spin frequency and pulsed flux
history of A 0535+26, which shows a mixed type I and II behavior with a series of normal outburst leading to a giant outburst which is
still in progress. In these and the following figures the r.m.s. pulsed flux is used.

\subsection{Disk-Fed Supergiants}

\begin{figure}[t!]
\centering
\includegraphics[width=3.2in]{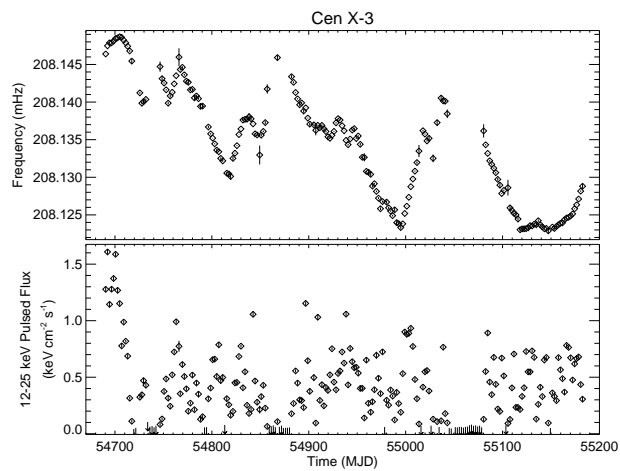}
\caption{Frequency and pulsed flux history for the eclipsing disk-fed supergiant pulsar Cen X-3 ($P_{orbit} = 2.08$\,d , $P_{spin}=4.80$\,s).} \label{cenx3}
\end{figure}

We detect one system, Cen X-3, with a supergiant companion that is overflowing, 
or nearly overflowing its Roche-lobe, resulting in a persistent accretion disk. The GBM spin frequency and pulsed flux history for Cen X-3 is shown
in Figure \ref{cenx3}. Cen X-3, an eclipsing binary, shows a remarkable ``torque switching'' behavior with persistent spin-up alternating with intervals of persistent spin-down,
with relatively short transitions between the two states. The torque behavior is not simply correlated with the flux.

\subsection{Wind-Fed Supergiants} 
\begin{figure}[!b]
\centering
\includegraphics[width=3.2in]{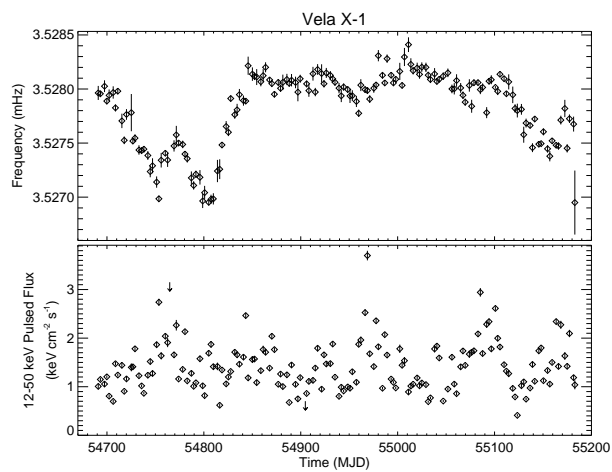}
\caption{Frequency and pulsed flux history for the eclipsing wind-fed supergiant pulsar Vela X-1 ($P_{orbit} = 8.96$\,d , $P_{spin}= 283$\,s).} \label{velax1}
\end{figure}

\begin{figure}[!t]
\centering
\includegraphics[width=3.2in]{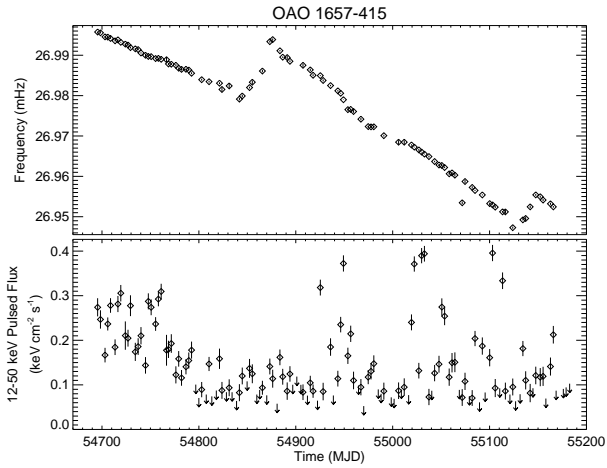}
\caption{Frequency and pulsed flux history for the eclipsing wind-fed supergiant pulsar OAO 1657-415 ($P_{orbit} = 1.70$\,d , $P_{spin}=31.24$\,s).} \label{oao1657}
\end{figure}

We have detected four systems with supergiant companions where Roche-lobe is underfilled, within the accretion flow from the
companion in the form of a stellar wind. These are Vela X-1, GX 301-2, 4U 1538-52, and OAO 1657-415. In Figure \ref{velax1} we show the GBM spin frequency 
and pulsed flux of Vela X-1. This source, which is an eclipsing binary, is the prototypical wind accretor. The frequency history is consistent with a
random walk, or white torque noise. In Figure \ref{oao1657} we show the GBM spin frequency and pulsed flux of OAO 1657-415, also an eclipsing binary.
In contrast to Vela X-1, it shows torque switching, suggesting that an accretion disk which is stable for months forms from the companions wind.

\subsection{Persistent LMXBs.} 

\begin{figure}[!b]
\centering
\includegraphics[width=3.2in]{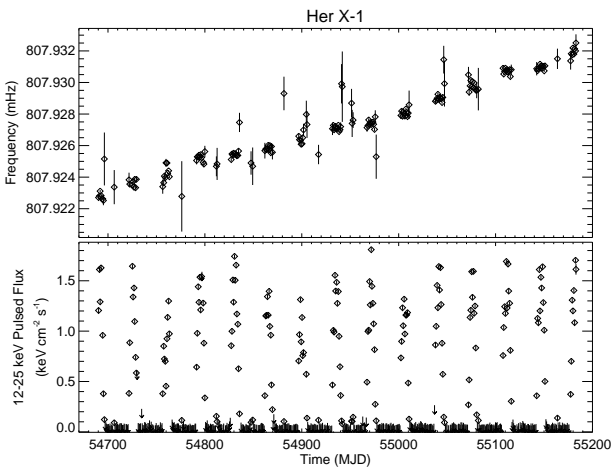}
\caption{Frequency and pulsed flux history for the eclipsing LMXB pulsar Her X-1 ($P_{orbit} = 1161$\,d , $P_{spin}=155$\,s).} \label{herx1}
\end{figure}

\begin{figure}[!b]
\centering
\includegraphics[width=3.2in]{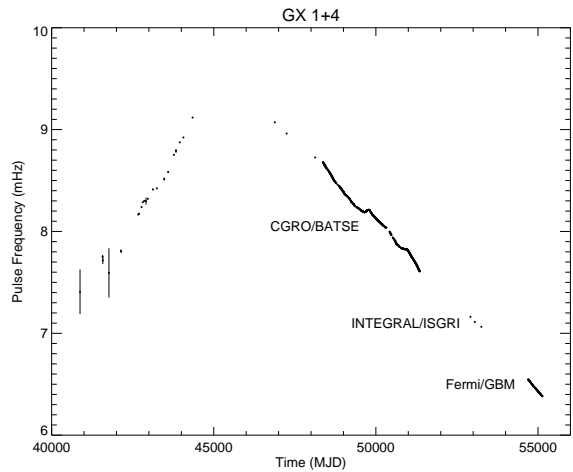}
\caption{The long-term pulse frequency history for GX 1+4.} \label{gx1p4_2}
\end{figure}

We have detected three accreting pulsars with low mass companions, 4U 1626-67, Her X-1, and GX 1+4. 
The characteristics of each are unique. 4U 1626-67 is an ultra-compact binary with the companion overflowing its Roche-lobe resulting in an accretion disk. 
For the GBM observations of
this source see A. Camero-Arranz et al. in these proceedings \cite{Camero-Arranz_2009}. Her X-1 is an eclipsing Roche-lobe overflow system. Obscuration of the pulsar by the accretion disk
results in a 35 day flux cycle of a ``Main-on'' state, off state, ``Short-on" state, and another off state. The spin-frequency and pulsed flux history is
shown in Figure \ref{herx1}. GX 1+4 is a symbiotic binary, with the pulsar in a wide orbit accreting from the wind of its red giant companion. The GBM pulse frequency
and pulsed flux history are shown in Figure \ref{gx1p4}. These observations show persistent spin-down, with the spin-down rate having no clear relationship to
the pulsed flux. The long-term history of the GX 1+4 pulse frequency is shown in Figure \ref{gx1p4_2}, which shows a change from spin-up to spin-down in 1983,
with only two short returns to spin-up since. 

\section{Conclusion}
\begin{figure}[!t]
\centering
\includegraphics[width=3.2in]{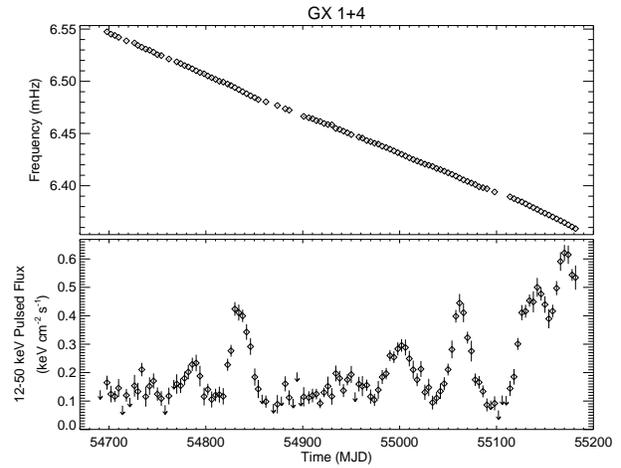}
\caption{Frequency and pulsed flux history for persistent LMXB (and symbiotic binary) pulsar GX 1+4 ($P_{orbit} = 1161$\,d , $P_{spin}=155$\,s).} \label{gx1p4}
\end{figure}

We have established a pulsed source monitoring system which uses the Fermi GBM data to rapidly detect previously
unknown or newly active accreting pulsars, and track the behavior of all detectable pulsars. This system has two components: 1) a daily blind search for pulsed sources, 
and 2) a monitor of known sources, which uses narrow band pulse searches specific to each source. To date we have detected 17 accreting pulsars. For each of the
sources detected we are producing long term histories of the pulse frequence, flux and pulse profile in three energy channels. The plots in Figures \ref{exo2030} to
\ref{gx1p4} are taken from these history files.  Updates of these figures and those for other sources may be found at 
http://gammaray.nsstc.nasa.gov/gbm/science/pulsars/\,. We are currently monitoring five sources that we have yet to detect, and will be regularly increasing the
number of sources in our monitoring catalog.

\begin{acknowledgments}
M.H.F. acknowledges support from NASA grant NNX08AG12G.
\end{acknowledgments}

\bigskip 

\end{document}